%% file: main.tex
\documentclass[sigconf, natbib, screen=true, anonymous=false, review=false]{acmart}

\acmSubmissionID{2201}

\input{packages}
\input{definitions}
\input{meta}

\input{authors}

\makeatletter
\gdef\@copyrightpermission{
\begin{minipage}{0.3\columnwidth}
\href{https://creativecommons.org/licenses/by/4.0/}
{\includegraphics[width=0.90\textwidth]{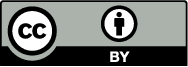}}
\end{minipage}\hfill
\begin{minipage}{0.7\columnwidth}
\href{https://creativecommons.org/licenses/by/4.0/}
{This work is licensed under a Creative Commons Attribution International 4.0 License.}
\end{minipage}
\vspace{5pt}
}
\makeatother

\begin{document}

\title{Predictive Uncertainty-based Bias Mitigation in Ranking}

\begin{abstract}

Societal biases that are contained in retrieved documents have received increased interest. Such biases, which are often prevalent in the training data and learned by the model, can cause societal harms, by misrepresenting certain groups, and by enforcing stereotypes. 
Mitigating such biases demands algorithms that balance the trade-off between maximized utility for the user with fairness objectives, which incentivize unbiased rankings.
Prior work on bias mitigation often assumes that ranking scores, which correspond to the utility that a document holds for a user, can be accurately determined. 
In reality, there is always a degree of uncertainty in the estimate of expected document utility. 
This uncertainty can be approximated by viewing ranking models through a Bayesian perspective, where the standard deterministic score becomes a distribution. 

In this work, we investigate whether uncertainty estimates can be used to decrease the amount of bias in the ranked results, while minimizing loss in measured utility.
We introduce a simple method that uses the uncertainty of the ranking scores for an uncertainty-aware, post hoc approach to bias mitigation. We compare our proposed method with existing baselines for bias mitigation with respect to the utility-fairness trade-off, the controllability of methods, and computational costs. We show that an uncertainty-based approach can provide an intuitive and flexible trade-off that outperforms all baselines without additional training requirements, allowing for the post hoc use of this approach on top of arbitrary retrieval models.

\end{abstract}


\maketitle

\acresetall
\input{sections/introduction.tex}
\input{sections/related_work.tex}
\input{sections/method.tex}
\input{sections/experiments.tex}
\input{sections/results.tex}
\input{sections/discussion.tex}

\input{sections/conclusion.tex}

\header{Data and code}
To facilitate reproducibility of our work, all code and parameters are shared at \url{https://github.com/MariaHeuss/2023-CIKM-uncertainty-based-bias-mitigation}.

\begin{acks}
    The research was partially funded by the Hybrid Intelligence Center, a 10-year program funded by the Dutch Ministry of Education, Culture and Science through the Netherlands Organisation for Scientific Research, \url{https://hybrid-intelligence-centre.nl}, and project LESSEN with project number NWA.1389.20.183 of the research program NWA ORC 2020/21, which is (partly) financed by the Dutch Research Council (NWO).
    All content represents the opinion of the authors, which is not necessarily shared or endorsed by their respective employers and/or sponsors.
\end{acks}

\clearpage

\bibliographystyle{ACM-Reference-Format}
\balance
\bibliography{references}
\clearpage 

\end{document}

%% file: packages.tex
\usepackage{amsmath}
\usepackage{amsfonts}
\usepackage{dsfont}
\usepackage{mathtools}
\usepackage{graphicx}
\usepackage{booktabs} 
\usepackage{numprint}
\npthousandsep{,}
\npdecimalsign{.}
\usepackage[inline]{enumitem}
\usepackage{balance}
\usepackage[skip=0pt]{caption}
\usepackage{multirow}
\usepackage{siunitx}
\usepackage{acronym}
\usepackage{xcolor}
\usepackage{algorithm}
\usepackage{algpseudocode}
\usepackage{caption}
\usepackage{subcaption}

\usepackage{csvsimple}

\usepackage{pgfplots}
\usepackage{tikz}
\usetikzlibrary{calc}

\usepackage{xspace}

%% file: definitions.tex
\AtBeginDocument{%
  \providecommand\BibTeX{{%
    \normalfont B\kern-0.5em{\scshape i\kern-0.25em b}\kern-0.8em\TeX}}}

\newcommand{\headernodot}[1]{\vspace{1mm}\noindent\textbf{#1}}
\newcommand{\header}[1]{\headernodot{#1.}}

\newcommand\restr[2]{{
  \left.\kern-\nulldelimiterspace 
  #1 
  \vphantom{\big|} 
  \right|_{#2} 
  }}

\acrodef{IR}{information retrieval}

\newcommand{\method}{PUFR\xspace}
\newcommand{\longmethod}{\textbf{P}redictive \textbf{U}ncertainty based
\textbf{F}air \textbf{R}anking\xspace}
\newcommand{\data}{\mathcal{D}}

\newcommand{\MAP}{\text{MAP}}
\newcommand{\map}{\text{MAP}}

\allowdisplaybreaks

%% file: meta.tex
 \providecommand\BibTeX{{%
  Bib\TeX}}

\copyrightyear{2023}
\acmYear{2023}
\setcopyright{rightsretained}
\acmConference[CIKM '23]{Proceedings of the 32nd ACM International Conference on Information and Knowledge Management}{October 21--25, 2023}{Birmingham, United Kingdom}
\acmBooktitle{Proceedings of the 32nd ACM International Conference on Information and Knowledge Management (CIKM '23), October 21--25, 2023, Birmingham, United Kingdom}\acmDOI{10.1145/3583780.3615011}
\acmISBN{979-8-4007-0124-5/23/10}

\ccsdesc[500]{Information systems~Retrieval models and ranking}
\ccsdesc[500]{Information systems~Evaluation of retrieval results}

%% file: authors.tex
\author{%
Maria Heuss%
}
\orcid{0000-0001-5360-9627}
\affiliation{%
 \institution{
 University of Amsterdam
 \city{Amsterdam}
 \country{The Netherlands}%
 }  
}  
\email{m.c.heuss@uva.nl}

\author{%
Daniel Cohen%
}
\orcid{0000-0002-2724-8247}
\affiliation{%
 \institution{Dataminr
 \city{NYC}
 \country{USA}%
 }  
}  
\email{daniel.cohen@dataminr.com}

\author{%
Masoud Mansoury%
}
\orcid{0000-0002-9938-0212}
\affiliation{%
\institution{
 University of Amsterdam}
 \institution{
 Discovery Lab, Elsevier}
 \city{Amsterdam}
 \country{The Netherlands}
}  
\email{m.mansoury@uva.nl}

\author{%
Maarten de Rijke%
}
\orcid{0000-0002-1086-0202}
\affiliation{%
\institution{
 University of Amsterdam
 \city{Amsterdam}
 \country{The Netherlands}%
 }  
}  
\email{m.derijke@uva.nl}

\author{%
Carsten Eickhoff%
}
\orcid{0000-0001-9895-4061}
\affiliation{%
 \institution{
 University of T\"ubingen 
 \city{T\"ubingen }
 \country{Germany}%
 }  
}  
\email{c.eickhoff@acm.org}

%% file: sections/introduction.tex
\section{Introduction}

The probability ranking principle (PRP)~\citep{robertson1977probability} states that, for optimal retrieval, the documents should be ranked in order of the predicted probability of relevance to the user. 
While this principle is ideal with respect to user utility, a ranking approach that solely relies on this principle can lead to an unfair treatment of the documents through unfair exposure and learned historical biases that are implicit in the data~\cite{biega2018equity,zehlike2017fa}. 
This realization has led to a broad range of work in the field of \emph{fair} ranking, where ways of ranking are explored that do not always strictly follow the PRP, but instead correct for such historical biases and distribute exposure more fairly~\cite{castillo2019fairness, ekstrand2022overview, patro_fair_ranking_survey}.
Such biases can be reflected in different ways, e.g., models can be biased to over proportionally favor members of one group over another~\cite{angwin2016machine}.
In this work, we follow \citet{rekabsaz2021societal}, 
and say that a ranking model is \emph{biased}, if documents that contain biases or stereotypes towards a protected group, e.g., people identifying with a certain gender, are being placed in ranked lists for queries that should be inherently neutral.

\header{Using uncertainty to mitigate biases and improve fairness}
Recent work has highlighted how learned ranking models violate the PRP -- that each score is not well calibrated, and that learned ranking models do not provide an equally reliable estimate of a document's relevance~\cite{penha2021uncertDropout, cohen2022downstreamuncertainty}. 
In this work, we take advantage of this violation of assumptions to produce a fair ranking with minimal utility loss. 
Rather than relying on a deterministic score, we consider the \textit{uncertainty of the model's estimate} to violate the PRP in an informed manner by focusing on the most \textit{uncertain documents}.

Our proposed method, called \longmethod (\method) exploits  knowledge about the certainty of the predicted relevance scores for mitigating bias by intervening at the scoring distribution, making it a post-processing method that is easy to use on top of arbitrary ranking models. 
Furthermore, \method does not require any training or fine-tuning of supervised models. Rather, given a ranked list of documents generated by a ranking model (most likely biased), \method leverages the uncertainty of the predicted scores assigned to the candidate documents by the ranking model to modify the ranked list among the most uncertain positions to generate a fairer ranking.  \method aims to reduce the impact of biased documents, while adhering to the PRP as closely as possible, only intervening in places where the ranking model was not very certain to begin with.

Additionally, we introduce an entirely post hoc uncertainty quantification procedure, based on Laplace approximation, that allows \method to approximate the uncertainty for any off the shelf model without access to the training data or optimization procedure. This is in contrast to past work that requires a specific training regime to produce the uncertainty scores for each candidate~\cite{penha2021uncertDropout, cohen2021not, cohen2022downstreamuncertainty, yang2022marginalcertainty}.

\header{Motivating example} In Fig.~\ref{fig:visualization}, we visualize our approach to predictive uncertainty-based fairness, \method. In this example, the objective is to promote the unbiased documents (marked in green) to appear on top of the ranked result.
We start by considering not only the mean ranking score but also the score distribution (uncertainty) as visualized with the cross resp.\ curve in Fig.~\ref{fig:visualization-a}. 
We chose confidence intervals relative to the standard deviation in which we allow \method to adjust the scores for each document, as can be seen in Fig.~\ref{fig:visualization-b}. 
Depending on whether a document is biased or not, we increase the score in this confidence interval if the document is unbiased or decrease it otherwise as visualized with the green/red crosses in Fig.~\ref{fig:visualization-c}. As the confidence intervals of the second (D2) and third (D3) documents \textit{intersect}, this changes the order of these scores. 
After re-ranking with respect to the newly obtained scores, the protected document D3 has swapped place with the non-protected document D2 as seen in Fig~\ref{fig:visualization-d}. 
As there are minimal computational costs for \method, developers/users have the freedom to
modify 
the trade-off between utility and fairness with minimal costs 
for their
use-cases. 

\input{figures/visualisation.tex}

\header{Our contributions} 
We summarize our contributions as follows: 
\begin{itemize}[leftmargin=*,nosep]
\item We introduce the notion of uncertainty-based fair ranking and analyze the potential of using the model uncertainty w.r.t.\ the ranking scores for bias mitigation. 
\item We define \method, an intuitive re-ranking approach that takes as input the ranking score distribution and calculates new ranking scores that can be used to create a less biased ranked list, while still preserving some certainty guarantees.
\item We compare \method to several in- and post-processing bias mitigation methods and show that it outperforms all baselines, while being computationally much less expensive than some of them. Moreover, we demonstrate that \method is easily controllable with respect to the trade-off between fairness and utility, making it practical for use in real-life ranking applications. 
\end{itemize}

%% file: figures/visualisation.tex
\begin{figure}[t!]

\begin{subfigure}{.45\linewidth}
  \includegraphics[width=\linewidth]{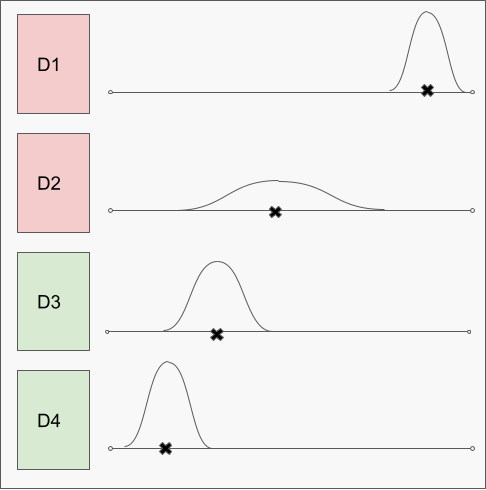}
  \vspace*{-5mm}
  \caption{}
  \label{fig:visualization-a}
\end{subfigure}\hfill 
\begin{subfigure}{.45\linewidth}
  \includegraphics[width=\linewidth]{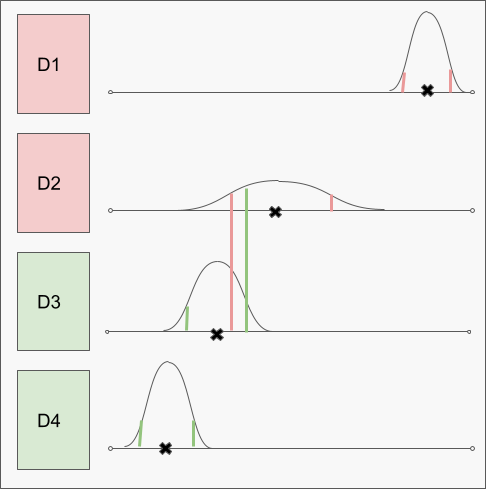}
  \vspace*{-5mm}
  \caption{}
  \label{fig:visualization-b}
\end{subfigure}

\medskip 
\begin{subfigure}{.45\linewidth}
  \includegraphics[width=\linewidth]{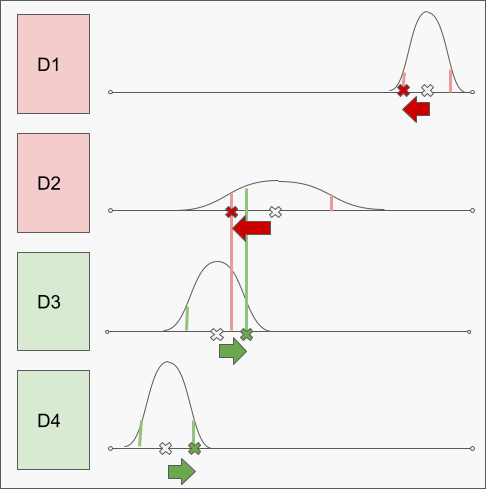}
  \vspace*{-5mm}
  \caption{}
  \label{fig:visualization-c}
\end{subfigure}\hfill 
\begin{subfigure}{.45\linewidth}
  \includegraphics[width=\linewidth]{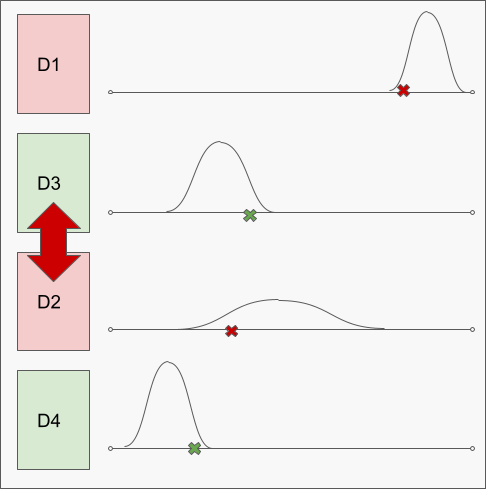}
  \vspace*{-5mm}
  \caption{}
  \label{fig:visualization-d}
\end{subfigure}
\caption{Visualization of our method \method. Next to the mean ranking scores \method also considers the score distribution that we obtained from the ranking model (\ref{fig:visualization-a}). Through intersecting confidence intervals (\ref{fig:visualization-b}) that allow us to adjust the scores (\ref{fig:visualization-c}) such that a not biased document, visualized in green, is swapping place with a higher ranked, biased document (\ref{fig:visualization-d}).}
\label{fig:visualization}
\vspace*{-3mm}
\end{figure}

%% file: sections/related_work.tex
\section{Related Work}\label{chap:Related-work}

\subsection{Uncertainty in ranking}
\citet{Zhu2009QLuncertainty} introduce the notion of considering a model's confidence when ranking documents. 
The authors view the confidence of a score based on the probabilistic model's own estimate -- the variance. 
Alternatively, we can assume a Bayesian perspective that considers how well the training data support the current model. 
As this approach does not rely on a probabilistic ranking model, it complements current ranking regimes.
\citet{penha2021uncertDropout} first introduce this notion of uncertainty into conversational retrieval by incorporating dropout into a BERT architecture at inference time. 
The ranking score is then modified by an uncertainty measure to improve the final re-ranking.
\citet{cohen2021not} suggest a similar approach for ad hoc retrieval where only the last layer's uncertainty is measured to offset both the complexity of a neural model and the size of the document set with similar re-ranking improvements. 
\citet{yang2022uncertnL2R} extend the above work by leveraging the uncertainty estimate to improve the exploration of an online learning to rank model. 
Rather than performing uncertainty-aware re-ranking, the uncertainty estimate is used to take an optimistic perspective on candidate documents to reduce the exploitation bias commonly found in an online learning to rank setting.

\subsection{Mitigating bias and fair ranking}

Recent years have seen a broad range of research on uncovering and mitigating biases in different information retrieval systems, such as biases in talent pool  \cite{geyik2019fairness} and resume search \cite{chen2018investigating} and the reinforcement of gender biases through search engines \cite{fabris2020gender}. 
\citet{rekabsaz2020neural} explore the extent to which  documents with gender bias can be found in the retrieved results of different neural retrieval models. 
Other work focuses more on the mitigation of such biases~ \citep[e.g.,][]{zerveas2022mitigating, rekabsaz2021societal}, where models are optimized to contain fewer biased documents for queries that are inherently unbiased. \citet{rekabsaz2021societal} use adversarial learning to remove gender bias from the trained model, \citet{zerveas2022mitigating} optimize the query representation from a previously trained architecture instead.

Mitigating biases is often framed as a fairness task. 
\citet{zehlike2022fairness, zehlike2022fairness2} introduce a classification framework for fair ranking approaches,  which we partly use to position our work in the existing fair ranking literature. 
As opposed to score-based fairness \citep{celis2020interventions,kleinberg2018selection,stoyanovich2018online,yang2019balanced}, where the ranking scores are assumed to be known, in this work we focus on supervised learning to rank, where the ranking scores need to be determined with a ranking model.

A large body of work focuses on \emph{merit-based} fairness, where the goal is to distribute the user attention in some way proportional to the merit of either individual documents (individual fairness~\citep[e.g.,][]{lahoti2019operationalizing, sarvi-2021-understanding-arxiv,heuss-2022-fairness}) or groups of documents~\citep[e.g.,][]{singh2018fairness, biega2018equity,wang2020fairness}. In contrast, other work \cite[e.g.,][]{zehlike2017fa, zehlike2022fairtopk} focuses on \emph{representational} fairness, which is concerned with removing historical biases from the ranking or representing documents from different groups fairly w.r.t.\ some demographic within the ranking. 

Independently of the notion of fairness, we differentiate between pre-processing~\cite{lahoti2019ifair}, in-processing~\cite{zehlike2017fa, singh2018fairness, biega2018equity, zehlike2020reducing,singh2019policy,beutel19, zerveas2022mitigating, rekabsaz2021societal}, and post-processing~\cite{zehlike2022fairtopk, diaz2020evaluating,kletti2022pareto} approaches to fairness interventions. These methods come into play either before the model is being trained, adjust the model or training process itself, or intervene after the model has been trained and the ranking scores are determined. 

\method is a \emph{post-processing} approach that aims to mitigate bias (\emph{representational} unfairness) as opposed to prior in-processing work on the same task \cite{rekabsaz2021societal, zerveas2022mitigating}. 
While other work on post-processing approaches \citep[such as, e.g.,][]{zehlike2017fa, celis2017ranking} intervene at the ranked output, our approach instead adjusts the score distribution.  
What distinguishes \method from prior work on fair ranking is that we aim to exploit the uncertainty that the ranking model has on the predicted relevance scores to increase the fairness of the rankings. 

\subsection{Uncertainty in fair ranking} 
 
Prior work at the intersection of uncertainty and fairness can be grouped into two categories.
The first category deals with uncertainty introduced when group membership cannot be determined with confidence.  
\citet{ghosh2021fair} discover that, when group labels are inferred from data,
the usage of fair ranking methods can invalidate fairness guarantees and even increase the disadvantage that protected groups might receive. 
\citet{mehrotra2022fair} follow up on this work and develop a fair ranking framework for cases where socially-salient group attributes cannot be determined with certainty but are assumed to follow a given probability distribution. 

The other category, which contains, among others, our work, considers the predictive uncertainty stemming from imperfect prediction of merits and ranking scores. 
\citet{yang2022marginalcertainty} are concerned with uncertainty in the relevance estimation. 
Unlike our work, the authors study an online setting where the relevance estimation is constantly updated. 
We target a static setting, not aiming to reduce the uncertainty for some exploration strategy but to exploit the uncertainty to obtain a better trade-off between fairness and utility.

Lastly, \citet{singh2021fairness} are concerned with uncertainty in merit due to observations of secondary attributes instead of directly observing the merit. 
The authors suggest a probabilistic fairness framework in the presence of such uncertainty. Their work defines a notion of fairness that takes the uncertainty in the merit prediction into account, while we exploit uncertainty to, for example, correct for historic biases in the data and ranking model.

\smallskip\noindent%
In summary, where existing methods either ignore the predictive uncertainty of ranking scores, aim to either reduce uncertainty, or take it into account when defining fairness, our work is the first to harness uncertainty to improve the fairness-utility trade-off. 

%% file: sections/method.tex
\section{Method}
\label{chap:Method}

We take an uncertainty-based approach to post hoc bias mitigation in ranking. We exploit the model's uncertainty over the predicted ranking scores to manipulate the ranking in a way that benefits documents that do not contain biases, which results in a fairer ranked list. By staying within a certain confidence range, we minimize the potential cost to utility. Following prior work~\cite{rekabsaz2021societal, mehrotra2022fair}, we frame the task as a fair ranking problem.

Our method operates entirely through principled machinery and allows us to trade-off between user utility and fairness by adjusting a single coefficient. Furthermore, an existing ranker can be used as-is, without the need to retrain it, making it possible to use and adjust it for various levels of fairness, with little additional costs. 

Below, in Section ~\ref{chap:notation}, we start by defining our notation and the fair ranking task. In Section~\ref{chap:Method-Risk-aware-fairness}, we introduce our method \method that, assuming that the predictive uncertainty over the ranking scores is given, uses those uncertainty values to develop a fair ranking approach. Finally, in Section~\ref{chap:Method-Uncertainty} we follow with a description of how to attain the uncertainty of a given deterministic ranking model over its scores at inference time.

\subsection{Notation and preliminaries}\label{chap:notation}
Given a query $q$, we consider the task of ranking documents from a candidate set $\mathcal{D}_q= \{d_{q,i}\}_i$ w.r.t.\ their relevance, to $q$.   
Regarding measured user utility only, an ideal ranked list would be ordered by decreasing document relevance.
We assume a ranking model has been trained to order the documents w.r.t.\ the relevance to the query by predicting relevance scores. 
Most rankers are deterministic, outputting only a single predicted relevance score, $\mu_{q,i}$. In Section~\ref{chap:Method-Uncertainty} we will describe how to approximate the uncertainty of predicted scores for such a model. 
We write $\sigma_{q,i}$ for the standard deviation of the predicted score $\mu_{q,i}$ for document $d_{q,i}$.  
Note that we implicitly assume the score distribution to be Gaussian.

Prior work has shown that models that are trained solely for maximizing the measured utility can be biased and contain unfair representations of the resulting ranked lists~\cite{rekabsaz2020neural}. In this work, as an additional objective, we aim to decrease the presence of biased documents in the ranked lists.  We treat the task as a fair ranking problem, where we want to increase the exposure of the protected group $\mathcal{D}_q^{P} \subset \mathcal{D}_q$ of documents without biases and decrease the exposure of the non-protected group $\mathcal{D}_q^{N} \subset \mathcal{D}_q$ of documents that contain biases.

\subsection{
\method: Uncertainty-aware fairness}
\label{chap:Method-Risk-aware-fairness}

In this section, we introduce our post-processing fairness intervention method \longmethod, \method. 
The core idea of \method is to take advantage of the uncertainty of the model over the predicted ranking scores to adjust these scores proportional to the standard deviation of the predictive distribution for each document, allowing fairness adjustments with minimal cost to the utility. 
For now, we treat the score distribution for each document, $\mathcal{N}(\mu_{q,i},\sigma^2_{q,i})$, as being given, but in Section~\ref{chap:Method-Uncertainty} we describe how to obtain it for a deterministic ranker. 

As the goal of \method is to mitigate bias and hence increase the fairness of the ranking system, \method accomplishes this by
swapping some of the documents of the protected group, $\mathcal{D}_q^{P}$, with higher ranked documents of the non-protected group, $\mathcal{D}_q^{N}$. 
Since the uncertainty of the scores for the documents within the same group can differ greatly, this allows for a tuned adjustment of the ranking scores where swaps only occur in settings where there exists a reasonable chance of the documents being equally relevant, quantified by the model's uncertainty, $\sigma_{q,i}$.

In other words, we allow \method to pick ranking scores that maximize fairness in intervals $[\mu_{q,i}-\alpha \cdot \sigma_{q,i}, \mu_{q,i}+\alpha \cdot \sigma_{q,i}]$, without re-ordering the documents within the same group. Here, $\alpha$ is a user defined hyper-parameter that quantifies the chance of a utility violation when performing this procedure.  
A higher value of $\alpha$ will result in a fairer ranking but at the cost of less accurate predicted scores, and hence potentially a drop in utility. 

As shown in Algorithm~\ref{alg:PUFR}, \method initially loops over all documents of the protected group $d_{q,i}\in\mathcal{D}_q^P$, sorted w.r.t.\ decreasing ranking score, $\mu_{q,i}$, see line~\ref{alg2:line2}. \method then increases the score as much as possible while staying within the confidence bounds, i.e.,
\begin{equation}
\tilde \mu_{q,i} = \mu_{q,i} + \alpha \cdot \sigma_{q,i}.
\end{equation}
See line~\ref{alg2:line3}. To avoid intra-group swapping of documents, modified ranking scores are bounded by the lowest score of any higher ranked document within the same group: 
\begin{equation}
\tilde \mu_{q,i} \leq \text{min}_{\mathcal{D}_q^P, j\leq i}(\tilde \mu_{q,j}),
\end{equation} 
where $j,i$ are rank positions, see line~\ref{alg2:line3b}.
Equivalently, for all documents of the non-protected group, $d_{q,i}\in\mathcal{D}_q^N$, we decrease the score as follows, this time starting with the document with the lowest ranking score (see line~\ref{alg2:line4}):
\begin{equation}
\tilde \mu_{q,i} = \mu_{q,i} - \alpha \cdot \sigma_{q,i},
\end{equation}
see line~\ref{alg2:line5}. Again, to avoid the same intra-group swapping for the non-protected group, we lower bound the adjusted scores by the maximum score of all documents in the same group that are ranked lower in the original ranking:
\begin{equation}
\tilde \mu_{q,i} \geq \text{max}_{\mathcal{D}_q^N, j\geq i}(\tilde \mu_{q,j}).
\end{equation} 
See line~\ref{alg2:line5b}.
\method then uses these adjusted scores $\tilde \mu_{q,i}$ to re-rank the documents (line~\ref{alg2:line8}).

Note that even though we define \method for a setting with only one protected document group, it can be extended to several protected groups, that need to receive different treatments. Our approach allows us to adjust the strength of the score adjustment individually for each group, e.g., enabling a stronger correction for more disadvantaged groups, by allowing a group-wise choice of hyper-parameter $\alpha_g$.

\newcommand{\mutil}{\tilde{\mu}}

\begin{algorithm}[t]
\caption{\longmethod (\method)}
\label{alg:PUFR}
\begin{algorithmic}[1]
\Require mean ranking scores $\{\mu_{q,i}\}_{d_{q,i}\in \mathcal{D}_q}$, 
standard deviation $\{\sigma_{q,i}\}_{d_{q,i}\in \mathcal{D}_q}$, control parameter $\alpha$, groups $\mathcal{D}_q^P$, $\mathcal{D}_q^N$ 

\ForAll{$d_{q,i}\in \mathcal{D}^P_q$, sorted by decreasing $\mu_{q,i}$}\label{alg2:line2}
        \State $\mutil_{q,i} \gets \mu_{q,i} + \alpha \cdot \sigma_{q,i}$\label{alg2:line3}
        \State $\mutil_{q,i} \gets \text{max}_{\mathcal{D}_q^P, j\leq i}(\mutil_{q,j})$\label{alg2:line3b}
\EndFor
\ForAll{$d_{q,i}\in \mathcal{D}^N_q$, sorted by increasing $\mu_{q,i}$}\label{alg2:line4}
        \State $\mutil_{q,i} \gets \mu_{q,i} - \alpha \cdot \sigma_{q,i}$\label{alg2:line5}
        \State $\mutil_{q,i} \gets \text{min}_{\mathcal{D}_q^N, j \geq i}(\mutil_{q,j})$\label{alg2:line5b}
\EndFor
\State Obtain ranking \textbf{$L$} by sorting documents ${d_{q,i}\in \mathcal{D}_q}$ with respect to scores $\mutil_{q,i}$\label{alg2:line8} 
\State \textbf{return} $L$
\end{algorithmic}
\end{algorithm}

Many pre-trained ranking models do not output the uncertainty scores $\sigma_{q,i}$ that \method employs to reorder rankings. Thus we need a way to approximate the uncertainty scores $\sigma_{q,i}$ in a post-processing manner. Next, we show how to do this with the help of Laplace approximation.

\subsection{Attaining uncertainty scores from a deterministic ranking model} 
\label{chap:Method-Uncertainty}

The goal is to attain effective uncertainty scores, $\sigma$, from a ranking model at inference time; conventional uncertainty approaches fail to satisfy this condition~\cite{penha2021uncertDropout,cohen2021not,yang2022uncertnL2R, yang2022marginalcertainty}. Past approaches have relied on a specific training regime -- Monte Carlo (MC) dropout -- to achieve an effective Bayesian model. As \method is a post hoc method, we leverage an alternative form of uncertainty, \textit{Laplace approximation}, that can be applied to any already trained ranking model.

The standard approach to training a deterministic model $f$, where there exists a single output for each input, is to learn a set of parameters, $\theta_{\map}$, that minimizes the loss function 
\begin{equation}
    \label{eq:loss_map}
    \mathcal{L}(\theta) = -\ln P(\theta\mid\mathcal{D}) + r(\theta),
\end{equation}
where  $r$ is some regularization on $\theta$ and $\data$ is the training dataset.
While this is a probabilistic interpretation of the loss function and optimization process, prior work has mapped margin-based ranking losses to this framework~\cite{cohen2021not}. 
At inference time, the model, $f$, is evaluated using the single point $\theta_{\map}$, which minimizes $\mathcal{L}(\theta)$. Alternatively, a Bayesian perspective captures the uncertainty of the model by considering all possible $\theta$ values weighed by how likely they are based on the training data using the posterior $P(\theta\mid \data)$, with $\theta_\map$ as the most likely value. This produces a distribution over outputs, of which the variance $\sigma^2$ represents the uncertainty present within the model and $\data$:
\begin{equation}
\label{eq:pred_dist}
    P(y\mid x,\mathcal{D}) = \int_\theta P(y\mid x, \theta)P(\theta\mid \mathcal{D}) d\theta, 
\end{equation}
with $x$ as the input and $y$ as the output of the model. Unfortunately, capturing this distribution is intractable for all but the smallest models due to the nature of computing the posterior $P(\theta\mid \mathcal{D})$. 
There exists prior work that approximates this distribution using MC Dropout~\cite{penha2021uncertDropout,cohen2021not,yang2022uncertnL2R, yang2022marginalcertainty}. However, this requires a specific training regime, which would prevent the general application of \method to off-the-shelf architectures or previously trained ranking models. 

\header{Using Laplace approximation for post-hoc uncertainty approximation} We propose using Laplace approximations (LA), which can turn any conventionally trained deterministic model into a Bayesian model at inference time to produce the necessary $\sigma$ values for \method~\cite{mackay1992approxBayesian}.  LA encompass a family of approaches that fit a local Gaussian around the MAP estimate (\ref{eq:loss_map}) via a second-order Taylor expansion of the log posterior:
\begin{equation}
\begin{aligned}
    \label{eq:laplace}
        \ln P(\theta\mid \data) \approx{} & \ln P(\theta_{\MAP}\mid \theta)
    \\ 
        & \frac{1}{2}(\theta-\theta_{\MAP})^\intercal \bar{H} (\theta-\theta_{\MAP}),
\end{aligned}    
\end{equation}
where $\bar{H}$ is the expected Hessian at $\theta_{\MAP}$.  The key observation is that the right side only requires the deterministic model, $\theta_{\MAP}$ to produce the log Bayesian posterior distribution on the left side.
Then, to recover the full posterior, exponentiating both sides reveals the Gaussian functional form for $\theta$,
\begin{equation}
\begin{aligned}
    \label{eq:laplace_gaussian}
    P(\theta\mid \data) \approx {}& P(\theta_{\MAP}\mid\data) - \\ 
        & \exp\left({\frac{1}{2}(\theta-\theta_{\MAP})^\intercal \bar{H} (\theta-\theta_{\MAP})}\right) \\
        {}\approx {}& \mathcal{N}(\theta_{\MAP}, \bar{H}^{-1}).
\end{aligned}
\end{equation}
Thus, this approximation can take any twice differentiable off-the-shelf model and conveniently convert it to a Bayesian model at inference time by inverting the Hessian. While inverting to produce the covariance matrix is intractable for most models, we leverage past work by only inverting the last layers of a neural model to achieve actionable uncertainty estimates with near-zero cost~\cite{cohen2022downstreamuncertainty, cohen2021not} (Algorithm \ref{alg:LLA}, lines~\ref{alg1:line2}--\ref{alg1:line3}).
While there exists a closed form linearization of Eq.~\ref{eq:laplace_gaussian}, we are able to achieve sufficient efficiency using Monte Carlo sampling to capture the predictive distribution $P(y\mid x,f)$ by sampling from the Gaussian (line \ref{alg1:line5}),  $\mathcal{N}(\theta_{\MAP}, \bar{H}^{-1})$~\cite{deng2022accelerated},
\begin{equation}
\begin{aligned}
    \label{eq:mc_predictive}
    P(y\mid x,\mathcal{D}) = &\int_{\theta} P(y\mid x,\theta)  P(\theta\mid \data)  d \theta  \\ 
    \approx & \frac{1}{N} \sum_{t=1}^{N}  p(y\mid x,\theta_t), \theta_t \sim \mathcal{N}(\theta_{\MAP}, \bar{H}^{-1})   .
\end{aligned}
\end{equation}
Furthermore, as the covariance matrix $H^{-1}$ is viewed as independent to the training process, we do not need to use the original loss function either~\cite{kristiadi2020lastlayerbayesian}. Lastly, for further efficiency, we exploit the property that the Hessian, $H$, is equivalent to the Fisher information matrix, $F$, at $\theta_{\map}$. As shown in Algorithm~\ref{alg:LLA}, we therefore approximate $H$ by taking the diagonal of $F$, which is a common approximation regime (line~\ref{alg1:line3})~\cite{gorantla2021problem,ritter2018kfacLA}.

After estimating  $\mathcal{N}(\theta_{\MAP}, \bar{H}^{-1})$ for the last layer of a neural model,  we sample this distribution $N$ times to produce $N$ versions of the last layer, in order to produce  $\mu_{q,\cdot}$ and $\sigma^2_{q,\cdot}$ as parameters of the predictive distribution $P(y\mid x,\mathcal{D}) = \mathcal{N}(\mu_{q,\cdot}, \sigma^2_{q,\cdot})$ (line \ref{alg1:line7}--\ref{alg1:line8}). These parameters are then used by \method as described in Section~\ref{chap:Method-Risk-aware-fairness} to debias the ranked list.  

\begin{algorithm}[t]
\caption{Post hoc uncertainty estimation for single query}
\label{alg:LLA}
\begin{algorithmic}[1]
\Require pre-trained $l$-layer model $f_\theta$, $\theta_{\MAP}=[\theta^1_{\MAP}, \ldots, \theta^l_{\MAP}]$, query $q$, candidate documents $\mathcal{D}_q = \{d_{q,i}\}_i$, Monte Carlo sample size $N$.
\ForAll{$d_{q,i}\in \mathcal{D}_q$}
    \State $h^{l-1}_i,y = f_{\theta_{\map}}(q,d_{q,i})$ \label{alg1:line2}
    \State $H \approx \text{diag}(F) = \text{diag}(
\mathbb{E}\big[\nabla_{\theta^l} \ln P(y\mid q,d_{q,i}))^2 \big]$ \label{alg1:line3}
    \ForAll{$j \in N$}
        \State $\{\theta\}_1^j \sim \mathcal{N}({\theta^l}_{\map}, \text{diag}F^{-1})$ \label{alg1:line5}
    \EndFor
        \State $\mu_{q,i} = \frac{1}{N} \sum_{t=1}^{N}  f_{\theta_t^l}(h_i^{l-1})$ \label{alg1:line7}
        \State $\sigma^2_{q,i} =  \frac{1}{N} \sum_{t=1}^{N}  f_{\theta_t^l}(h_i^{l-1})^2 - \left( \frac{1}{N} \sum_{t=1}^{N}  f_{\theta_t^l}(h_i^{l-1}) \right)^2$ \label{alg1:line8}
        
\EndFor
\State \textbf{return} $\mu_{q,i}, \sigma_{q,i}$ $\forall d_{q,i} \in \mathcal{D}_q$
\end{algorithmic}
\end{algorithm}

%% file: sections/experiments.tex
\section{Experimental Setup}\label{chap:experimental-setup}

We aim to answer the following research questions with our experiments: 
\begin{enumerate*}[label=(RQ\arabic*)]
    \item Based on empirical findings, are the uncertainty intervals around the ranking scores of a Bayesian ranking model sufficiently intersecting to allow for a re-ranking of documents, while staying within reasonable certainty bounds? \label{RQ:uncertainty_intervals}    
    \item Can \method be used to reduce the number of biased documents that are ranked on top of the list more effectively than prior methods?\label{RQ:trade-off}    
    \item How do the various methods for fairness interventions compare with respect to controllability and computational efficiency?\label{RQ:controllability}
\end{enumerate*}

There are four properties that we consider relevant to answer these questions:
\begin{enumerate*}[label=(\roman*)]
\item We want to improve the fairness within the rankings. 
\item We want to do so with the least loss in utility possible. 
\item The next property is the controllability of the approach at hand. A human user/engineer should be able to easily adjust the trade-off between fairness and utility to fit their purposes. 
\item The last property is computational efficiency since this can also play a role when choosing a fairness method.
\end{enumerate*}

Next, we detail our experimental design. 
Then we discuss the evaluation metrics that we use to measure the four properties mentioned above (Section \ref{sec:experiments-eval}) and the dataset that we use (Section~\ref{sec:experiments-datasets}).
Section~\ref{sec:experiments-baselines} summarizes the baselines that we compare against. 

\subsection{Experimental design}
\label{sec:experiments-design} 
We perform our experiments on a web search task, where for each query, the objective is to rank documents that might be relevant to that query. 
In addition to the requirement of being relevant to the user, the ranked list should not contain any gender biases for queries that are naturally non-gendered~\cite{rekabsaz2021societal}. 
Therefore, we consider only non-gendered queries and expect a fair ranking model to not promote any documents that are biased towards some gender.
See Section~\ref{sec:experiments-datasets} for a discussion on the data used for this task. 

To get an effective impression of the trade-off between utility and fairness, we perform a range of experiments per baseline, by varying some hyperparameter $\alpha$. 
We define this hyperparameter individually for each baseline, based on the respective underlying algorithms (see Section~\ref{sec:experiments-baselines}).  

To demonstrate the efficacy of \method on current search models, we use the BERT ranker introduced by~\citet{nogueira2019BERTreranker} as it represents a common language model architecture in current ranking regimes~\cite{santhanam2022colbertv2, hofstatter2021tasb, dong2022PLMexample,lin-2021-pretrained}. 
Due to hardware constraints, we use Bert-Mini \cite{turc2019miniBERT}, a distilled four-layer version of BERT that performs comparably to the full model in search and other related tasks. 
We note that in the case of uncertainty modeling, \citet{cohen2021not} demonstrate that a distilled model results in less expressive ranking uncertainty compared to larger variants of the same architecture on the same data. 
Thus, Bert-Mini represents a challenging setting and a conservative estimate of \method's performance.

To facilitate reproducibility of our work, all code and parameters are made available; see Section~\ref{chap:conclusion}.

\subsection{Evaluation}\label{sec:experiments-eval} 

User utility and fairness are measured per query. To get a single score to compare across methods, we report the mean over all queries. We measure significance with paired t-tests, where we treat the results of each query as one sample. 

\header{User utility} 
To measure user utility, we use the nDCG metric (normalized discounted cumulative gain). We use different cut-offs to measure the user utility in the top-$10$ documents, as well as for the first 100 documents. 

\header{Fairness} 
As discussed in Section~\ref{sec:experiments-design}, our task entails reducing the impact of strongly biased documents in the presented rankings. 
Therefore, we use the nFaiRR metric as a measure of fairness introduced by \citet{rekabsaz2021societal}. 
For a ranked list $L$, the FaiRR score at cut-off $k$ is defined as: 
\begin{equation}
\operatorname{FaiRR@k}(L)= \sum_{\operatorname{rank}_L(d_i)\le k} n_{d_i}\cdot \frac{1}{\operatorname{rank}_L(d_i)},
\end{equation}
where $\operatorname{rank}_L(d_i)$ describes the rank of candidate document $d_i$ in $L$, and the neutrality score $n_{d_i}\in [0,1]$ is lower, the more biased a document is. 
Since the possible range of FaiRR scores depends on the distribution of neutrality scores of its candidate documents, to make the results easier to interpret and better comparable among queries, we use the \emph{normalized FaiRR score} (nFaiRR).
For this, we normalize the FaiRR score with the highest attainable FaiRR score with the document candidates for this query, similar to how nDCG is calculated from DCG.
In our experiments we measure the nFaiRR at a cut-off value of 10 and 50. We select a different cut-off than the utility measure (@100) so as to compare with reported values from the baseline evaluations. 

\header{Controllability}
We follow prior work \citep{rekabsaz2021societal}, and focus on a qualitative analysis of the results by investigating the predictability of the utility-fairness trade-off when adjusting the controllable hyperparameter of each of the methods.  An ideal approach should have small change in utility and fairness for a small change in $\alpha$. 
To this end, we compare the plots in Fig.~\ref{figure:controllability} below.

\header{Computational efficiency}
For computational efficiency, we measure the run time of our implementation for each approach. 
We acknowledge that method-specific performance optimization might be able to further improve on the run times observed for the generic implementations used here, but assume that at least a rough execution time comparison can be gleaned. 
We measure the run time of each query and report the mean run time in Table~\ref{table:results}. 

\header{Significance testing} 
To test the significance of observed differences in evaluation scores, we perform two-tailed paired t-tests on the metrics, treating the results of an approach of each query as a measurement of the same random variable. 
In Table~\ref{table:results}, we mark results with an asterisk if they are significantly different from those of \method. 

\subsection{Dataset}
\label{sec:experiments-datasets} 
The retrieval models that we use are trained on the MS MARCO Passage Retrieval collection~\cite{nguyen2016ms}. 
For evaluation, we use $\text{MS MARCO}_\text{Fair}$, a subset consisting of 215 queries from the validation set that are non-gendered in nature -- i.e., not containing any words or concepts that could be attributed to some gender \cite{rekabsaz2021societal}. However, the top candidate documents for these queries are highly associated with gender~\cite{zerveas2022mitigating, rekabsaz2021societal}. We quantify the degree of gender bias for each document using the neutrality scores provided by \citet{rekabsaz2020neural} in order to measure fairness. We define documents with neutrality score 1 as the protected group for the post-processing baselines and \method.

\subsection{Baselines}
\label{sec:experiments-baselines}
The baseline fairness intervention methods that we consider include the two in-processing approaches that have been introduced for the same bias mitigation task and dataset used here \cite{rekabsaz2021societal, zerveas2022mitigating}. 
Since \method is a post-processing approach, we add two commonly used post-processing fairness approaches that have been slightly adjusted to fit the task. Both post-processing baselines as well as UNFAIR use the mean scores $\mu_{q,i}$, produced by Algorithm~\ref{alg:LLA} in  Section \ref{chap:Method-Uncertainty} for the BERT-based ranker (see Section \ref{sec:experiments-design}) as ranking scores. For each baseline the hyper-parameter $\alpha$ that allows us to control the trade-off between utility and fairness, is defined individually. 

\header{UNFAIR} The ranking resulting from ordering the documents with respect to the mean scores $\mu_{q,i}$, without considering fairness.

\header{ADV} The (in-processing) adversarial fairness optimization from~\citep{rekabsaz2021societal}, which shares the same underlying BERT re-ranking architecture as discussed in Section \ref{sec:experiments-design}. However, training is done using an adversarial discriminator head that attempts to predict whether the document is gendered or neutral by optimizing a classification loss function. 
The gradient from this loss is reversed within the main BERT architecture, therefore moving the parameters away from regions that can effectively capture gender \citep{ganin2015gradientreverse}. We implement this model using the source code and suggested hyperparameters provided by the authors. The controlling hyperparameter $\alpha$ (originally $\lambda$) is defined by the scale of the reversed gradient. 

\header{CODER} This (in-processing) baseline~\citep{zerveas2022mitigating} is intended for dense retrieval architectures. The method directly optimizes the query representation from a previously trained architecture, TAS-B~\cite{hofstatter2021tasb}, by jointly optimizing thousands of candidate documents in a list-wise manner. While improving overall ranking performance, the large candidate pool within a list-wise loss provides a stable and competitive way to incorporate fairness directly during training. We include this baseline not as a direct comparison with respect to ranking performance, but to provide context on how a direct list-based fairness optimization approach compares to methods that operate entirely within a post hoc framework when viewed from a utility-fairness trade-off perspective. Here, the hyperparameter $\alpha$ (in the original paper $\lambda_r$) is defined as the regularization coefficient for the neutrality loss.

\header{CVXOPT} A (post-processing) convex optimization approach similar to \cite{celis2017ranking}. For each query we optimize the ranking $L$ for utility, measured by nDCG, under a constraint on the nFaiRR score, $\operatorname{nFaiRR}(L)\ge \alpha$. To keep computational costs within a reasonable range, we only re-rank the first 50 documents of each query. 

\header{FA*IR} A (post-processing) approach suggested in \citep{zehlike2017fa}. We use a significance parameter $0.1$ as suggested in~\citep{zehlike2017fa} and vary $p$, the desired minimal proportions of documents with the protected attribute in the top-$k$ for any value of $k$. In the remainder of this paper we use $\alpha:=p$, not to be confused with the significance parameter in the original paper, to match the other methods. For a fair comparison w.r.t.\ to computational efficiency, we use an efficient implementation that pre-computes the required number of protected documents for each rank upfront via an iterative algorithm. 

%% file: sections/results.tex
\section{Experimental results}\label{chap:Results}

We present and discuss answers to our research questions.

\subsection{Intersections of uncertainty intervals}

Recall \ref{RQ:uncertainty_intervals}: \emph{Based on empirical findings, are the uncertainty intervals around the ranking scores of a Bayesian ranking model sufficiently intersecting to allow for a re-ranking of documents, while staying within reasonable certainty bounds?} 
To answer \ref{RQ:uncertainty_intervals}, we analyze the confidence intervals of the ranking scores. 
If the uncertainty intervals do not intersect much, the ranking model is very certain about the ordering of its ranking scores. 
In such a case, our approach, or any uncertainty-aware approach in general, would not be able to re-rank the documents within an acceptable utility bracket. 
Previous work has shown that ranking models tend to be very certain for the ranking scores of highly ranked documents~\cite{cohen2021not}, but certainty decays when going down the ranked list.  
We are interested in how much flexibility a rank-aware fairness approach would offer in swapping documents by allowing the ranking scores to take values in a given certainty $[\mu_{q, i}-\alpha\cdot \sigma_{q, i},\mu_{q, i}+\alpha\cdot \sigma_{q, i} ]$ interval around the mean score value $\mu_{q, i}$.  
Fig.~\ref{fig:median_swaps_1std_MSMARCO} shows the median number of documents with intersecting confidence intervals (i.e. the median number of documents that the document at that rank could swap position with) 
for $\alpha=1$ resp.\ $\alpha=2$ standard deviations.

\input{figures/median_swaps.tex}

Even for documents ranked at higher positions, there is flexibility to change the order of the ranking. For a confidence interval of 1 standard deviation, most documents in the top-10 each have at least 6 documents that they could swap rank with. 
If we look at confidence intervals of two standard deviations, this number increases to ${\sim}$10 documents that the document at rank 10 can swap place with. 
We therefore answer \ref{RQ:uncertainty_intervals} positively: The uncertainty intervals around the ranking scores of the Bayesian ranking model are sufficiently intersecting to allow for a re-ranking of documents, while staying within acceptable certainty bounds for utility. 

Having confirmed that within the uncertainty of the model there is  flexibility for an uncertainty-based fairness approach to change the order of documents, we address our second research question that asks whether the proposed approach can improve fairness. 

\subsection{The fairness utility trade-off}
Recall \ref{RQ:trade-off}: \emph{Can \method be used to reduce the number of biased documents that are ranked on top of the list more effectively than prior methods?}
To answer this question we refer to Fig.~\ref{fig:trade_off_utility_nfairr_10} and  \ref{fig:trade_off_utility_nfairr_50}, where we plot fairness on the x-axis against utility on the y-axis, for \method and the baselines  discussed in Section~\ref{sec:experiments-baselines}, for different values of the respective hyper-parameter $\alpha$ that controls the trade-off. In addition we use Table~\ref{table:results}, where we compare the experimental outcomes with the best nFaiRR value for a given minimum utility requirement.

\input{figures/utility-nfairr_plots.tex}

\header{Utility-fairness trade-off} 
In Fig.~\ref{fig:trade_off_utility_nfairr_10} and~\ref{fig:trade_off_utility_nfairr_50}, we observe that the CODER baseline starts with a better trade-off for the top-10 documents, which can be attributed to better ranking scores that it starts out with (\method uses a BERT-based model to obtain ranking scores). CODER's advantage quickly vanishes as the balancing parameter $\alpha$ increases for more weight on fairness. 
Overall, \method offers a better trade-off between fairness and utility than the CODER based and the adversarial fairness optimization baseline (ADV).

If we compare \method to the post-processing baselines (CVXOPT and FA*IR), it clearly outperforms those baselines. Once a nFaiRR value of 0.96 is reached the advantage of \method over these baselines becomes smaller. For a possible explanation see Section \ref{chap:discussion}.

\input{figures/results_tables.tex}

Overall, \method outperforms all baselines for a large range of nFaiRR values, which we also highlight by comparing the fairness of the different approaches at two different utility levels ($\operatorname{nDCG@100}=0.31$ and $\operatorname{nDCG@100}=0.30$) in Table~\ref{table:results}. 
We chose these levels of utility, assuming that, when taking a fair ranking approach in production there might be a certain (small) allowance for a drop in utility given, within which the best possible fairness value should be reached. 
We see that for these levels \method reaches significantly higher scores for nFaiRR than all baselines.

\header{Ablation study} To ensure that the uncertainty estimates indeed do contribute to the success of \method, we conduct an ablation study. We compare \method with a similar approach that, instead of adjusting the scores relative to the standard deviation, in- or decreases all scores by the same, constant value. In our experiments we use the mean uncertainty score over all queries and candidates documents, $\sigma_{\text{mean}}=\text{mean}_{q,i}(\sigma_{q,i})$. 
The results of this ablation study are presented in Fig.~\ref{fig:ablation_MSMARCO_top10}. We see that by using the uncertainty scores instead of a uniform correction factor, we gain a better trade-off. 
For the top-10, these improvements are less visible (see Fig.~\ref{fig:ablation_MSMARCO_top10} (a)).  
When considering the top-100 documents instead, the advantages of using uncertainty become much clearer (see Fig.~\ref{fig:ablation_MSMARCO_top10} (b)).
This might be due to fact that, as also noted by \citet{cohen2021not}, for the top-10 documents the uncertainty scores tend to be fairly similar to each other, making our approach, if we only look at a small window, seem similar to the ablation study approach. When we look at a larger window, the uncertainty scores deviate more, emphasizing the advantages of \method. 

We conclude this section and answer \ref{RQ:trade-off} in the affirmative. \method performs competitively with baselines. In terms of fairness-utility trade-offs it significantly outperforms other post-processing schemes, and clearly beats the two state-of-the-art in-processing baselines. The ablation study confirms that this result is at least partially due to the use of the model's uncertainty in its scores. Hence, \method can be used to reduce the number of biased documents that are ranked on top of the list more effectively than prior methods. 

Since a good utility-fairness trade-off is not the only relevant criterion when choosing a fair ranking method, our next research question \ref{RQ:controllability} concerns the degree of controllability and computational costs of the different methods.  

\input{figures/ablation-plot.tex}

\subsection{Controllability and computational efficiency}
Next, we address \ref{RQ:controllability}: \emph{How do the various methods compare with respect to controllability and computational efficiency?}
As discussed in Section \ref{sec:experiments-eval}, we focus on a qualitative  analysis of the $\alpha$-fairness and $\alpha$-utility curves, evaluating how predictable and hence controllable the utility-fairness trade-off is.  Fig.~\ref{figure:controllability} shows that for \method the nFaiRR score monotonically increases with increasing $\alpha$. At the same time, utility, measured by nDCG, decreases. Both curves are highly predictable. Furthermore, since re-ranking is computationally very efficient, a broad range of rankings with different trade-offs can be explored to find the right choice of hyper-parameter for the desired trade-off between nFaiRR and nDCG. 
The CODER-based approach has similarly predictable trade-off curves as \method~ \cite{zerveas2022mitigating}. 
However, CODER is an in-processing approach, meaning that the model needs to be re-trained for each choice of hyper-parameter $\alpha$, making it much less controllable in practice. 
The ADV method on the other hand, seems to be highly unpredictable, on top of the downsides that come with in-processing methods as discussed above. 
For the FA*IR baseline, although its curve seems to be fairly well controllable, the granularity in which we can produce results is much coarser. Due to space constraints we omit the figure for the convex optimization approach; because of computational efficiency, FA*IR or \method should be preferred over it. 

\input{figures/controllability_figures.tex}

With regard to computational efficiency, we recall that both in-processing approaches, ADV and CODER, once trained, do not have the post-processing overhead of the other methods. 
However, these methods need a large amount of training to gain a reasonable level of performance~\cite{rekabsaz2021societal, zerveas2022mitigating}. Looking at Table~\ref{table:results}, re-ranking with \method is much faster than with the other two post-processing approaches. Obtaining uncertainty labels can be done within microseconds. After adjusting the ranking scores there is a single re-sorting of the documents that dominates the execution time. Hence, when using \method in production and adjusting the score before the initial ordering of the documents, the execution of \method is nearly free. 

%% file: figures/median_swaps.tex

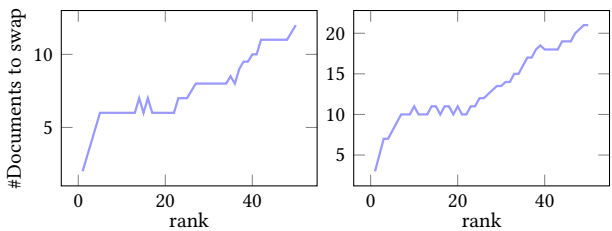
\begin{figure}
\centering

\newcommand{\SpacingX}{0.1em}
\newcommand{\Width}{0.28\textwidth}
\newcommand{\Height}{0.22\textwidth}
\newcommand{\BarWidth}{0.001\textwidth}
\newcommand{\BarOffset}{0.00\textwidth}
	\begin{tikzpicture}[font=\small]
    		\begin{axis}[
            	xlabel = rank,
			ylabel=\#Documents to swap,
            	width=\Width,
            	height=\Height,
            	y label style={yshift=-2em},
            	x label style={yshift=0.9em},
            	bar width=\BarWidth,
        	]       
        	\addplot[line width=0.3mm,color=blue!40!white] table[x=index, y=median_docs_to_swap, col sep=comma]{figures/csv_files/Median_num_docs_to_swap_1std_FAIRMARCO.csv}; 
    		\end{axis}
	\end{tikzpicture}%
	\begin{tikzpicture}[font=\small]
    		\begin{axis}[
            	xlabel = rank,
            	width=\Width,
            	height=\Height,
            	x label style={yshift=0.9em},
            	bar width=\BarWidth,
			legend style={nodes={scale=0.8, transform shape}},
        	]       
        	\addplot[line width=0.3mm,color=blue!40!white] table[x=index, y=median_docs_to_swap, col sep=comma]{figures/csv_files/Median_num_docs_to_swap_2std_FAIRMARCO.csv}; 
    		\end{axis}
    		legend style={at={(0.9,0.5)},anchor=west}
	\end{tikzpicture}%
	\caption{$\text{MSMARCO}_{\text{FAIR}}$: Median number of documents that have intersecting uncertainty intervals with the document placed at each rank for uncertainty intervals of 1 (left) resp. 2 (right) standard deviations. }\label{fig:median_swaps_1std_MSMARCO}
\end{figure}

%% file: figures/utility-nfairr_plots.tex
\begin{figure}
\newcommand{\Width}{0.95\columnwidth}
\newcommand{\Height}{0.65\columnwidth}

\centering
\begin{tikzpicture}
\begin{axis}[
            xlabel=nFaiRR@10,
			ylabel=nDCG@10,
            width=\Width,
            height=\Height,   
			legend pos=north west,
            legend cell align={left},
            legend style={fill opacity=0.75, draw opacity=1,text opacity=1},
            x label style={yshift=0.5em},  
            y label style={yshift=-0.5em},  
        	]       
\addplot[
        scatter,only marks,scatter src=explicit symbolic,
        scatter/classes={
            ours_adapted={mark=square*,fill=blue!40!white},
            adv={mark=square*,fill=yellow!40!white}, 
            coder={mark=diamond*,fill=purple!40!white}, 
            cvxopt={mark=triangle*,fill=green!40!white},
            fa*ir={mark=*,fill=orange!40!white}
        }
    ]
table [x=nfairr10, y=ndcg_cut_10, col sep=comma,meta=approach] {figures/csv_files/msmarco_baselines_10.csv};
\legend{\method, ADV, CODER, CVXOPT, FA*IR}
\end{axis}
\end{tikzpicture}
\caption{Trade-off between fairness and utility evaluated on the first 10 documents.} \label{fig:trade_off_utility_nfairr_10}
\end{figure}
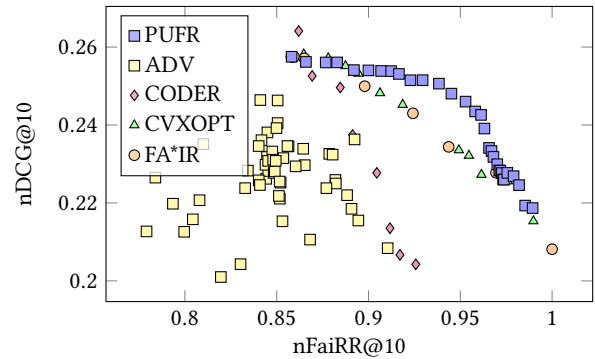

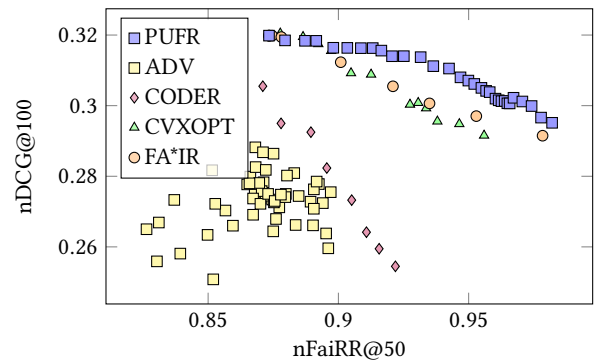
\begin{figure}
\newcommand{\Width}{0.95\columnwidth}
\newcommand{\Height}{0.65\columnwidth}
\centering
\begin{tikzpicture}
\begin{axis}[
            xlabel=nFaiRR@50,
			ylabel=nDCG@100,
            width=\Width,
            height=\Height,      
			legend pos=north west,
            legend cell align={left},
            legend style={fill opacity=0.75, draw opacity=1,text opacity=1},
            x label style={yshift=0.5em},  
            y label style={yshift=-0.5em}, 
        	]       
\addplot[
        scatter,only marks,scatter src=explicit symbolic,
        scatter/classes={
            ours_adapted={mark=square*,fill=blue!40!white},
            adv={mark=square*,fill=yellow!40!white}, 
            coder={mark=diamond*,fill=purple!40!white}, 
            cvxopt={mark=triangle*,fill=green!40!white},
            fa*ir={mark=*,fill=orange!40!white}
        }
    ]
table [x=nfairr50, y=ndcg_cut_100, col sep=comma,meta=approach] {figures/csv_files/msmarco_baselines.csv};
\legend{\method, ADV, CODER, CVXOPT, FA*IR}
\end{axis}
\end{tikzpicture}
\caption{Trade-off between fairness and utility evaluated on the first 50 resp. 100 documents.}
\label{fig:trade_off_utility_nfairr_50}
\end{figure}

%% file: figures/results_tables.tex
\begin{table}[t]
 \caption{Results for experiment with best nFairr value for nDCG decrease not more than 0.01 and 0.02 respectively. ADV baseline does not fulfill the criteria  of being at most 0.01 nDCG points worse than UNFAIR. * denotes significance w.r.t. \method via two tailed paired students t-test of $p<.05$.}
\setlength{\tabcolsep}{3pt}
\centering
  \begin{tabular}{l l c cc c c c c}    
  \toprule
	 & &
    	 &
        \multicolumn{2}{c}{nDCG{$\uparrow$}} &
        \multicolumn{2}{c}{nFaiRR{$\uparrow$}} &
        \multicolumn{1}{c}{re-rank-} & 
        req.
          \\
        \cmidrule(r){4-5}
        \cmidrule(r){6-7}
        &
        \multicolumn{1}{c}{Method} &
        \multicolumn{1}{c}{$\alpha$} &
        \multicolumn{1}{c}{@10} &
        \multicolumn{1}{c}{@100} &
        \multicolumn{1}{c}{@10} &
        \multicolumn{1}{c}{@50} &
        \multicolumn{1}{c}{time(s){$\downarrow$}} &
        train\\
        \midrule 
        &UNFAIR&0.0&0.26&0.32&0.858&0.873&0.00 & No\\
        &ADV&2.0&0.21&0.26&0.91&0.896& - & Yes\\
        \midrule
        \multirow{4}{*}{\rotatebox[origin=c]{90}{\scriptsize{\textbf{$nDCG_{100}\ge 0.31$}}}} 
        &\textbf{\method}&2.5&\textbf{\textbf{0.25}}&\textbf{0.31}&\textbf{0.938}&\textbf{0.932}&0.014&No\\
        
        &CODER&3.0&\textbf{0.25}&\textbf{0.31}&0.920\rlap{*}&0.920\rlap{*}& - & Yes\\
        &CVXOPT&0.8&\textbf{0.25}&\textbf{0.31}&0.906\rlap{*}&0.905\rlap{*}&0.123&No\\
        &FA*IR&0.7&\textbf{0.25}&\textbf{0.31}&0.898\rlap{*}&0.901\rlap{*}&0.058&No\\
        \midrule
        \multirow{4}{*}{\rotatebox[origin=c]{90}{\scriptsize{\textbf{$nDCG_{100}\ge 0.30$}}}} 
        
        &\textbf{\method}&7.0&0.23&\textbf{0.30}&\textbf{0.970}&\textbf{0.960}&0.014&No\\
        
        &CODER&4.0&\textbf{0.24}&\textbf{0.30}&0.927\rlap{*}&0.926\rlap{*}& - & Yes\\
        &CVXOPT&0.91&0.23&\textbf{0.30}&0.949\rlap{*}&0.931\rlap{*}&0.123&No\\
        &FA*IR&0.85&0.23&\textbf{0.30}&0.944\rlap{*}&0.935\rlap{*}&0.058&No\\
        \bottomrule
  \end{tabular}
 \label{table:results}
\end{table}

%% file: figures/ablation-plot.tex
\begin{figure}[!t]
\centering

\newcommand{\SpacingX}{0.5em}
\newcommand{\SpacingY}{-1.3em}
\newcommand{\Width}{0.26\textwidth}
\newcommand{\Height}{0.24\textwidth}
\newcommand{\BarWidth}{0.001\textwidth}
\newcommand{\BarOffset}{0.00\textwidth}
\begin{tikzpicture}
\begin{axis}[
            xlabel=nFaiRR@10,
			ylabel=nDCG@10,
			legend pos=south west,
            width=\Width,
            height=\Height,
            x label style={yshift=\SpacingX},
            y label style={yshift=\SpacingY},
            legend cell align={left}
        	]       
\addplot[
        scatter,only marks,scatter src=explicit symbolic,
        scatter/classes={
            ours_adapted={mark=square*,fill=blue!40!white},
            ablation={mark=*,fill=red!40!white}
        }
    ]
table [x=nfairr10, y=ndcg_cut_10, col sep=comma,meta=approach] {figures/csv_files/msmarco_ablation_10.csv};
\legend{\method,Ablation}
\end{axis}
\end{tikzpicture}
\centering
\begin{tikzpicture}
\begin{axis}[
            xlabel=nFaiRR@100,
			ylabel=nDCG@100,
			legend pos=south west,
            width=\Width,
            height=\Height,
            x label style={yshift=\SpacingX},
            y label style={yshift=\SpacingY},
            legend cell align={left}
        	]       
\addplot[
        scatter,only marks,scatter src=explicit symbolic,
        scatter/classes={
            ours_adapted={mark=square*,fill=blue!40!white},
            ablation={mark=*,fill=red!40!white}
        }
    ]
table [x=nfairr100, y=ndcg_cut_100, col sep=comma,meta=approach] {figures/csv_files/msmarco_ablation.csv};
\legend{\method,Ablation}
\end{axis}
\end{tikzpicture}
\caption{Ablation study comparing \method (score adjustment proportional to the ranker's uncertainty) with an ablation experiment with uniform score adjustment.}
\label{fig:ablation_MSMARCO_top10}
\end{figure}
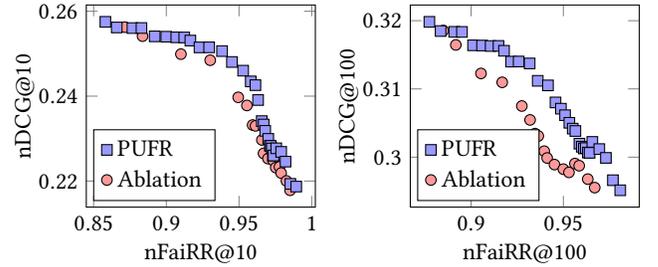

%% file: figures/controllability_figures.tex
\begin{figure*}\label{fig:trade_off_utility_nfairr}
\centering

\newcommand{\SpacingX}{0.1em}
\newcommand{\Width}{0.245\textwidth}
\newcommand{\Height}{0.21\textwidth}
\newcommand{\BarWidth}{0.001\textwidth}
\newcommand{\BarOffset}{0.00\textwidth}
\begin{tikzpicture}[font=\small]
    		\begin{axis}[
            	xlabel =alpha,
			    %
                ylabel =, 
            	width=\Width,
            	height=\Height,
            	y label style={yshift=-2em},
            	x label style={yshift=0.9em},
            	bar width=\BarWidth,
            	axis y line=right,
    			ymin=0.78,
    			ymax=1.02,
                yticklabels=\empty
        	]       
        	\addplot[line width=0.3mm,color=orange!40!white] table[x=alpha, y=nfairr10, col sep=comma]{figures/csv_files/ours_results.csv};
    		\end{axis}
    		\begin{axis}[
            	xlabel =,
			    ylabel=\ref{ndcg} nDCG@10,
            	width=\Width,
            	height=\Height,
            	y label style={yshift=-1em},
            	x label style={yshift=0.9em},
            	bar width=\BarWidth,
            	axis y line=left,
    			ymin=0.14,
    			ymax=0.27	
        	]       
    		\addplot[line width=0.3mm,color=teal!40!white] table[x=alpha, y=ndcg_cut_10, col sep=comma]{figures/csv_files/ours_results.csv};\label{ndcg}
    		\coordinate (legend) at (axis description cs:1.04,0.88);
    		\end{axis} 
            \node[above right] at (current bounding box.north) {\method};
	\end{tikzpicture}%
	\begin{tikzpicture}[font=\small]
    		\begin{axis}[
            	xlabel =alpha,
			    %
                ylabel=,
            	width=\Width,
            	height=\Height,
            	y label style={yshift=-2em},
            	x label style={yshift=0.9em},
            	bar width=\BarWidth,
            	axis y line=right,
    			ymin=0.78,
    			ymax=1.02,
                yticklabels=\empty
        	]       
        	\addplot[line width=0.3mm,color=orange!40!white] table[x=alpha, y=nfairr10, col sep=comma]{figures/csv_files/adv_results.csv};
    		\end{axis}
    		\begin{axis}[
            	xlabel =,
                ylabel=,
            	width=\Width,
            	height=\Height,
            	y label style={yshift=-2em},
            	x label style={yshift=0.9em},
            	bar width=\BarWidth,
            	axis y line=left,
    			ymin=0.14,
    			ymax=0.27,
                yticklabels=\empty	
        	]       
    		\addplot[line width=0.3mm,color=teal!40!white] table[x=alpha, y=ndcg_cut_10, col sep=comma]{figures/csv_files/adv_results.csv};
    		\coordinate (legend) at (axis description cs:1.04,0.88);
            \legend{};
    		\end{axis}    
            \node[above] at (current bounding box.north) {ADV};
	\end{tikzpicture}
	\begin{tikzpicture}[font=\small]
    		\begin{axis}[
            	xlabel =alpha,
                ylabel=,
            	width=\Width,
            	height=\Height,
            	y label style={yshift=-2em},
            	x label style={yshift=0.9em},
            	bar width=\BarWidth,
            	axis y line=right,
    			ymin=0.78,
    			ymax=1.02,
                yticklabels=\empty
        	]       
        	\addplot[line width=0.3mm,color=orange!40!white] table[x=alpha, y=nfairr10, col sep=comma]{figures/csv_files/coder_results.csv};
    		\end{axis}
    		\begin{axis}[
            	xlabel =,
                ylabel=,
            	width=\Width,
            	height=\Height,
            	y label style={yshift=-2em},
            	x label style={yshift=0.9em},
            	bar width=\BarWidth,
            	axis y line=left,
    			ymin=0.14,
    			ymax=0.27,
                yticklabels=\empty	
        	]       
    		\addplot[line width=0.3mm,color=teal!40!white] table[x=alpha, y=ndcg_cut_10, col sep=comma]{figures/csv_files/coder_results.csv};
    		\coordinate (legend) at (axis description cs:1.04,0.88);
    		\end{axis}    
            \node[above] at (current bounding box.north) {CODER};
	\end{tikzpicture}%
    \begin{tikzpicture}[font=\small]
    		\begin{axis}[
            	xlabel =alpha,
			    %
			    ylabel=\ref{nfairr} nFaiRR@10,
            	width=\Width,
            	height=\Height,
            	axis y line=right,
            	y label style={yshift=1em,rotate=-180},
            	x label style={yshift=0.9em},
            	bar width=\BarWidth,
    			ymin=0.78,
    			ymax=1.02,
        	]       
        	\addplot[line width=0.3mm,color=orange!40!white] table[x=alpha, y=nfairr10, col sep=comma]{figures/csv_files/fa_ir_results.csv};\label{nfairr}
    		\end{axis}
    		\begin{axis}[
            	xlabel =,
                ylabel=,
            	width=\Width,
            	height=\Height,
            	y label style={yshift=-2em},
            	x label style={yshift=0.9em},
            	bar width=\BarWidth,
            	axis y line=left,
    			ymin=0.14,
    			ymax=0.27,
                yticklabels=\empty
        	]       
    		\addplot[line width=0.3mm,color=teal!40!white] table[x=alpha, y=ndcg_cut_10, col sep=comma]{figures/csv_files/fa_ir_results.csv};\label{mrr}
    		\coordinate (legend) at (axis description cs:1.04,0.88);
    		\end{axis}    
            \node[above left] at (current bounding box.north) {FA*IR};
	\end{tikzpicture}%
	\caption{Controllability of different approaches visualized by plotting utility and fairness against the controlling hyper-parameter $\alpha$ on the x-axis (see Section \ref{sec:experiments-baselines} for a description of $\alpha$ for each approach).}
    \label{figure:controllability}
\end{figure*}
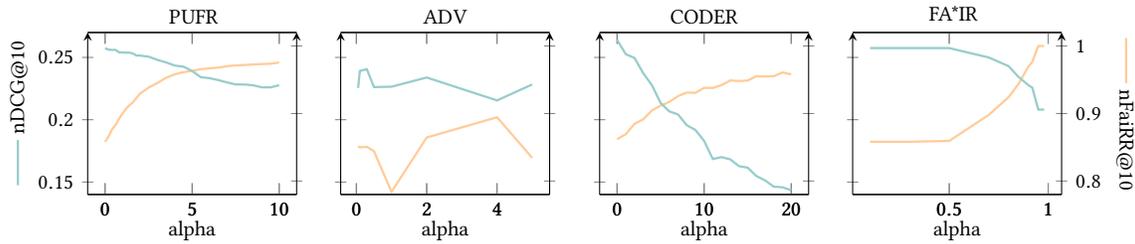

%% file: sections/discussion.tex
\section{Discussion}
\label{chap:discussion}

\begin{figure}[t!]
\begin{subfigure}{.475\linewidth}
  \includegraphics[width=\linewidth,  height=0.6\linewidth]{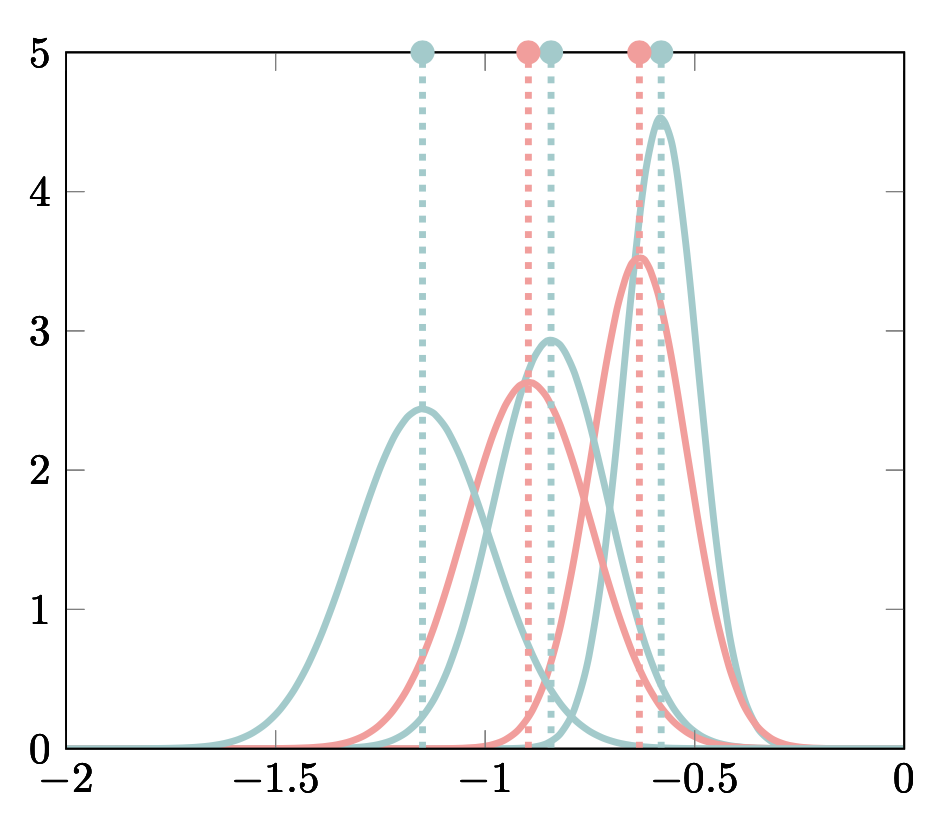}
  \caption{query 1032855 before \method}
  \label{figure:pufr-intersecting-before}
\end{subfigure}\hfill 
\begin{subfigure}{.475\linewidth}
  \includegraphics[width=\linewidth,  height=0.6\linewidth]{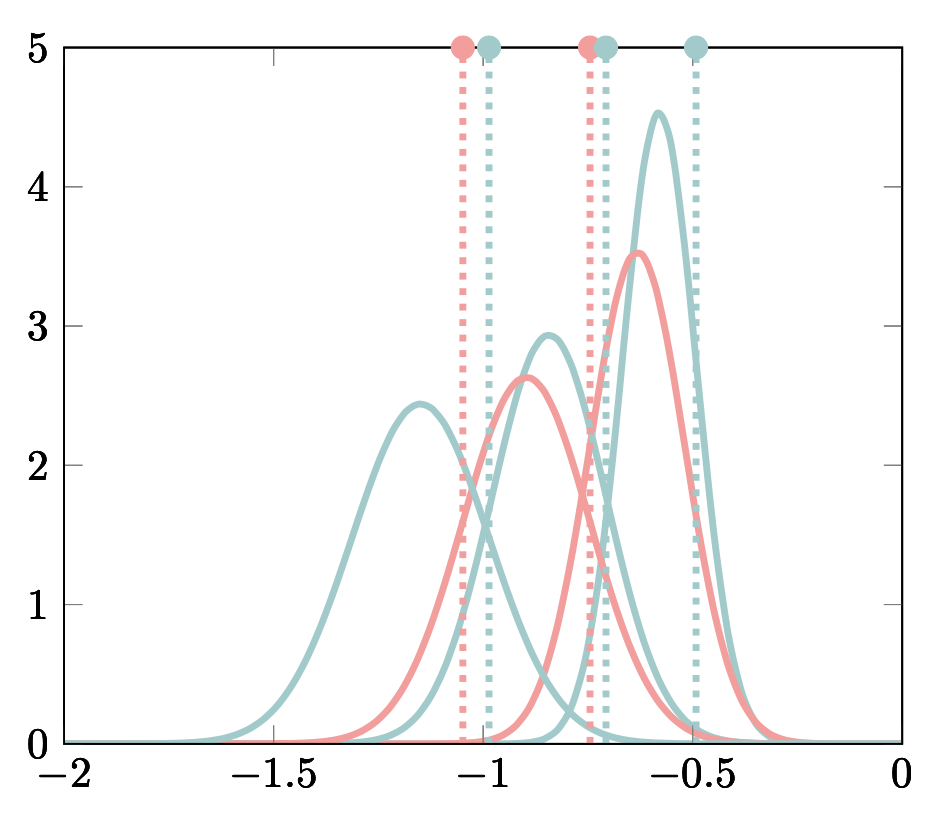}
  \caption{query 1032855 after \method}
  \label{figure:pufr-intersecting-after}
\end{subfigure}

\medskip 
\begin{subfigure}{.475\linewidth}
  \includegraphics[width=\linewidth,  height=0.6\linewidth]{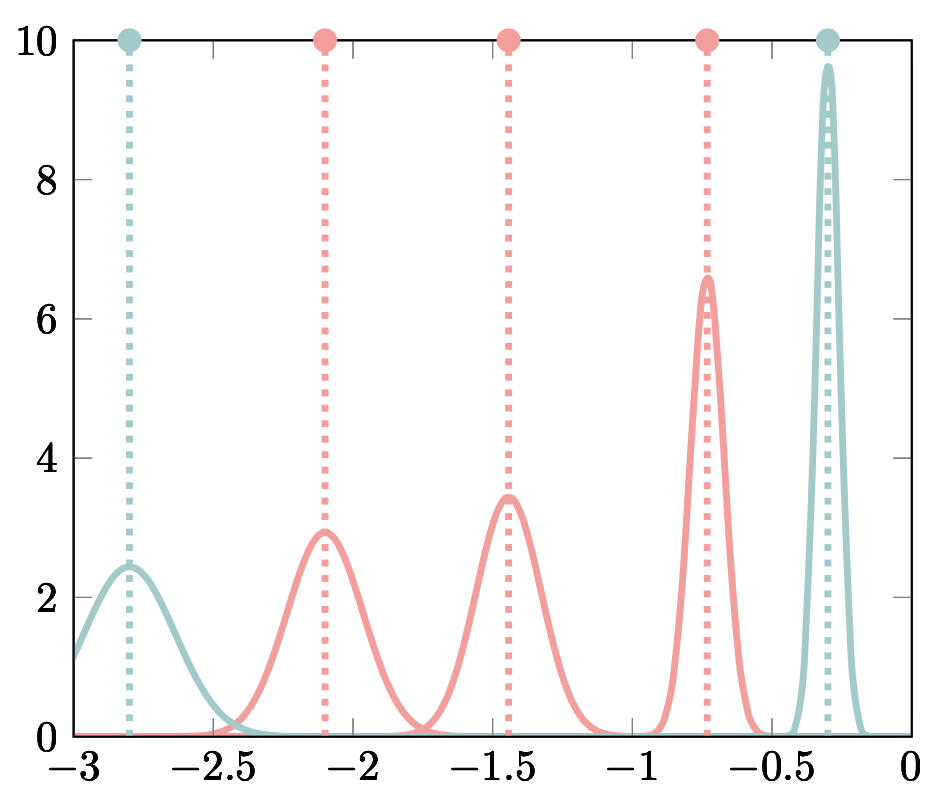}
  \caption{query 1089383 before \method}
  \label{figure:pufr-nonintersecting-before}
\end{subfigure}\hfill 
\begin{subfigure}{.475\linewidth}
  \includegraphics[width=\linewidth,  height=0.6\linewidth]{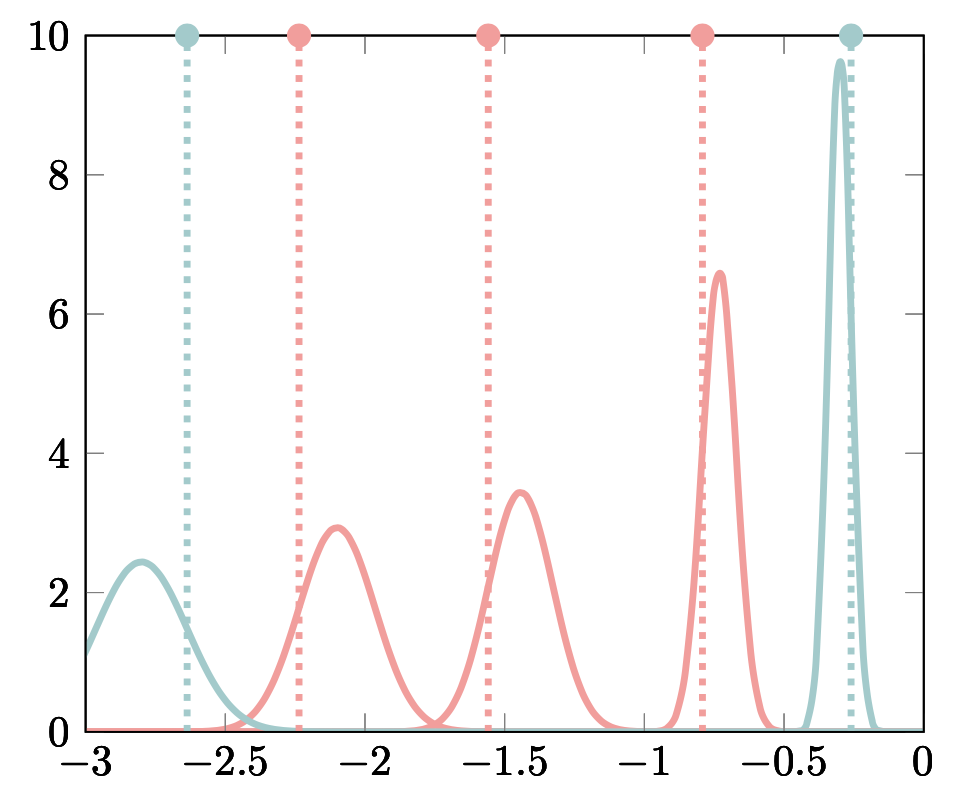}
  \caption{query 1089383 after \method}
  \label{figure:pufr-nonintersecting-after}
\end{subfigure}
\caption{Examples of score distributions for the top-5 documents for two queries of the MS MARCO$_{\text{Fair}}$ dataset. Protected documents in green, non-protected in red. Subfigs.~\ref{figure:pufr-intersecting-before} and~\ref{figure:pufr-nonintersecting-before} show the ranking score before \method adjusts the scores, ~\ref{figure:pufr-intersecting-after} and~\ref{figure:pufr-nonintersecting-after} show them after. Query 1089383 was scaled before plotting. }
\label{figure:before-after-pufr}
\end{figure}

\textbf{Exploiting model uncertainty for the fairness-utility trade-off.}
To increase the fairness of a ranking, we would commonly need to trade-off some predicted utility.Encouraging this trade-off to take place when the ranking model is less certain about the ranking scores will cause roughly equivalently relevant documents that the model cannot confidently rank, to swap place. Assuming that the ranking model is well calibrated, this might be the reason for the overall better trade-off that \method achieves, compared to models that do not consider predictive uncertainty. This quality is highlighted in Fig.~\ref{figure:before-after-pufr}, where we show the score distribution of the top-5 documents of two queries in the $\text{MSMARCO}_{\text{Fair}}$ dataset. 
In the case of Fig.~\ref{figure:pufr-intersecting-before} and \ref{figure:pufr-intersecting-after}, the larger variance leads to overlapping score distribution, allowing \method to swap documents in the re-ranked list. On the other hand, Fig.~\ref{figure:pufr-nonintersecting-before} and \ref{figure:pufr-nonintersecting-after} show a query where the model is very certain about the order of the documents. \method hence does not change the order of the documents, whereas FA*IR and CVXOPT both do adjust the ranking, leading to decreased user utility for those baselines.

\header{Using \method outside the models confidence}
Our empirical results show that if we allow \method to adjust the scores too far outside of its confidence, its performance starts to decay (see Fig.~\ref{fig:trade_off_utility_nfairr_10}). If $\alpha$ is too high, the natural interpretation of adjusting the scores within plausible error-bounds gets lost and we cannot exploit the models knowledge of its own certainty any further. Without the certainty to back it up, \method  becomes more arbitrary in its decisions where to trade-off predicted utility with fairness. 
Hence, \method is most effective for small values of $\alpha$, roughly up to $\alpha=4$ (see Fig.~\ref{figure:controllability}). 

This observation means that a purely uncertainty-based fairness method might not be the best choice when the bias we want to correct for is too strong. In such cases, it might be beneficial to use uncertainty in combination with another approach that has proven effective for the task at hand.

%% file: sections/conclusion.tex
\vspace*{-2mm}
\section{Conclusion}\label{chap:conclusion}

We have introduced the notion of predictive uncertainty-based ranking fairness, aiming to exploit a ranking model's uncertainty as an indicator of which documents we should focus on when re-ordering for a
fairer ranking which de-emphasizes documents containing biases. 
Through our empirical analysis we have found that the uncertainty intervals of the ranking scores are sufficiently intersecting to allow us to swap the position of some documents. We have also introduced an intuitive and principled post-processing method, \method, that adjusts the predicted ranking scores within some desired confidence bound. 
We have shown that by considering uncertainty,  \method can achieve the best utility-fairness trade-off and has superior time complexity and good controllability.

We hope that our contribution makes the adoption of methods to remove bias in ranked results more attractive to practitioners working on real- world search and recommendation systems.

More experimentation is needed to confirm our findings in more settings. 
We see limitations of our approach as twofold. 
Firstly, \method allows a re-ordering of the documents only within the uncertainty of the model. This might make our method less effective in reducing unfairness when the model is very skewed towards documents containing biases.
As a second limitation, we rely on uncertainty scores containing accurate information on which documents are more likely to be in the wrong order. Furthermore, the uncertainty intervals around the scores need to intersect sufficiently. In our experiments, we are using a neural ranking model on text data, which is a task that inherently carries a fair amount of uncertainty. For other tasks and fairness definitions, more research will be necessary to evaluate whether an uncertainty-based approach can be beneficial for the utility-fairness trade-off.

As to future work, an important next step would be to define ways to evaluate uncertainty scores in a listwise manner for ranking models. Without proper evaluation of the predictive uncertainty, we are unable to put trust on the score distribution and hence on an uncertainty-based fairness approach. 
Moreover, 
more work is needed to investigate whether \method could be extended to, for example, Bayesian learning-to-rank models or recommender systems. 
Finally, we see a clear need to create more datasets for large language models with fairness labels, on which methods such as ours can be tested.